\documentclass[fleqn,twoside]{article}
\usepackage{espcrc2}
\usepackage{graphicx}
\usepackage{amssymb}

\newcommand{\be}{\begin{equation}}
\newcommand{\ee}{\end{equation}}
\newcommand{\lap}{\lesssim}

\def\bea{\begin{eqnarray}}
\def\eea{\end{eqnarray}}
\def\la{\langle~}
\def\ra{~\rangle}
\def\dis{\displaystyle}

\title{QCD simulations at small chemical potential\thanks{Presented by T. Takaishi}}

\author
{Ph.\ de Forcrand\address{ETH-Z\"urich, CH-8093 Z\"urich and CERN Theory Division
, CH-1211 Geneva 23, Switzerland}, 
        S.~Kim\address{Dept. of Physics, Sejong University, Seoul 143-747, Korea} and
        T.~Takaishi\address{Hiroshima University of Economics, Hiroshima 731-0192, Japan}
}

\begin{document}
\begin{abstract}
Within the reweighting approach, one has the freedom to choose the Monte Carlo
action so that it provides a good overlap with the finite-$\mu$ measure 
but remains simple to simulate. We explore several choices of action in the
regime of small $\mu$. Simulating with a finite isospin chemical potential
$\mu_I=\mu$ gives a better overlap than the standard choice $\mu=0$,
with no computational overhead.
\end{abstract}

\maketitle

\section{INTRODUCTION}

Because heavy ion collision experiments
at RHIC  correspond to a chemical potential $\mu$ as small as 10MeV,
lattice QCD simulations at small $\mu$ have attracted renewed interest.
To address the sign problem caused by the complex fermion determinant,
three approaches have been proposed.
$(i)$ Analytically continuing results obtained at imaginary chemical 
potential~\cite{mu_I};
$(ii)$ Measuring the coefficients of the Taylor expansion of each observable
about $\mu=0$~\cite{GOT88,TARO};
$(iii)$ Measuring the observables from a reweighted ensemble~\cite{Glasgow}.
$(i)$ trades the sign problem for that of the analytic continuation; the 
approach works well for smooth functions of $\mu$~\cite{Owe-Ph}.
We will not discuss it further here.
$(ii)$ has no sign problem, but rapidly becomes complicated as the Taylor order increases.
$(iii)$ is enjoying a successful revival in a two-parameter version, where
results at $(\beta,\mu)$ are obtained by reweighting the $(\beta_0,\mu_0=0)$
simulation~\cite{FODOR,Bielefeld}.
The accessible range of chemical potentials and volumes 
is still limited by the sign problem. But also, ensuring a sufficient 
overlap between the desired ensemble at $(\beta,\mu)$ and the 
Monte Carlo ensemble at $(\beta_0,\mu_0=0)$ is notoriously difficult.
The acceptable shift of parameters normally decreases like 1/volume.     

It seems useful then to consider the sampling of an 
approximate finite-$\mu$ action. Reweighting to the exact action will be
mild and safe. The difficulty is to capture
most of the exact measure, while avoiding the computation of
the exact, complex determinant at each Monte Carlo step.
We explore three variants of this strategy below.

\section{SIMULATIONS AT SMALL $\mu$}

The QCD partition function with chemical potential, 
here for staggered quarks,
is given by 
\be
Z(\mu)=\int dU~\det M(\mu)^{\frac{n_f}{4}}~\exp(-S_G)
\label{eq1}
\ee
with $S_G$ the gluon action and $M(\mu)$ the Dirac \linebreak matrix.
Writing $\det M(\mu)~=~|\det M(\mu)|\exp(i\hat\theta)$
and defining $\theta \equiv \frac{n_f}{4} \hat\theta$,
Eq.(\ref{eq1}) becomes
\be
Z(\mu)=\int dU~|\det M(\mu)|^{\frac{n_f}{4}}e^{i\theta}~\exp(-S_G).
\ee
To allow Monte Carlo sampling over a positive measure, the expectation value
of an observable $O(\mu)$ is recast into
\be
\la O(\mu)\ra= \frac{\la O(\mu)~e^{i\theta} / f \ra_{\tilde{Z}}}
{\la e^{i\theta} / f \ra_{\tilde{Z}}}
\label{recast}
\ee
$\la .. \ra_{\tilde{Z}}$ means that the average is taken over
configurations sampled from the partition function~$\tilde{Z}$
\be 
\tilde{Z}=\int dU~|\det M(\mu)|^{\frac{n_f}{4}}~\exp(-S_G)~f(U)
\ee
where $f(U)$ is any positive functional of the gauge field. 
$e^{i\theta} / f$ is then the correction (reweighting) factor applied
to each configuration.
The standard
strategy corresponds to $\tilde{Z} = Z(\mu=0)$, i.e.   
\be
f~=~\left| \frac{\det M(\mu=0)}{\det M(\mu)}\right|^{\frac{n_f}{4}}
\label{f_standard}
\ee
One would like to find an optimal choice for $f(U)$. Setting aside the
algorithmic issue of sampling a given $f(U)$, the optimal choice is that
which maximizes the statistical accuracy on $\la O(\mu)\ra$ for an
ensemble of $N$ uncorrelated configurations. If $O(\mu)$
is not specified, it is reasonable to minimize the fluctuations in the
reweighting factor, i.e. the relative error on the
denominator of Eq.(\ref{recast}), since it propagates to all observables.
From the central limit theorem, that error is, for $N$ large
\be
\frac{1}{\sqrt{N}} \sqrt{\la (\frac{1}{f} \cos \theta)^2 \ra
/ \la \frac{1}{f} \cos \theta \ra^2 - 1}
\label{sd}
\ee
This expression is minimized for 
\be
f~=~| \cos \theta |
\label{f_opt}
\ee
which reduces the denominator of Eq.(\ref{recast}) to the average sign
$\la {\rm sgn} \cos \theta \ra_{\tilde{Z}}$.
Therefore, it would be desirable to sample from
\be
\tilde{Z}_{{\rm opt}}=\int dU~|\det M(\mu)|^{\frac{n_f}{4}}
~| \cos \theta |~\exp(-S_G)
\ee

The drawback of this sampling choice is that the phase $\theta$ of $\det M(\mu)$
must be evaluated at each Monte Carlo step, at a computational cost 
$\propto L^9$ for a lattice of size $L^3\times N_t$. Thus, this strategy seems
prohibitively inefficient.
We consider three alternatives, which can be implemented at reduced computing
cost and are closer to the optimal choice Eq.(\ref{f_opt}) than the standard
Eq.(\ref{f_standard}).

\noindent $\bullet$ {\bf Method A: $f = 1$} \\
The sampling measure is
$\propto\det^{\frac{n_f}{8}} [M^\dagger(\mu) M(\mu)]$.
Since $\det M^\dagger(\mu) = \det M(-\mu)$, this amounts to \linebreak 
simulating a finite {\em isospin} chemical potential~\cite{Stephanov} 
$\mu_I=\mu$.
Compared with the standard sampling \linebreak measure 
$\propto\det^{\frac{n_f}{4}} M(\mu=0)$,
this method captures, with no computational overhead, 
the fluctuations of the magnitude of the determinant
which may account for a good part of the relevant physics at small $\mu$.
Indeed, the critical line $T_c(\mu)$
in the isospin-$\mu$ case seems to have a similar curvature as in the usual 
isoscalar-$\mu$ case~\cite{Ejiri}.

\begin{table}[t]
\begin{center}
\begin{tabular}{|c|ccccc|}
\hline
$f(U)$ & $\mu=$ 0.1 & 0.15 & 0.20 & 0.25 & 0.30 \\
\hline
Eq.(\ref{f_standard}) & 0.089 & 0.22 & 0.45 & 0.97 & 2.67 \\
$\bf{1}$ & 0.085 & 0.17 & 0.33 & 1.08 & 3.83 \\
$e^{-\bar{\theta}^2/2}$ & 0.017 & 0.09 & 0.46 & 2.21 & 15.8 \\
$|\cos\bar{\theta}|$ & 0.008 & 0.05 & 0.46 & 14.4 & 14.2 \\
\hline
\end{tabular}
\end{center}
\caption{Standard deviation (numerator Eq.(\protect\ref{sd})) of the 
reweighting factor $\cos\theta / f$ as a function of $\mu$, 
for various sampling choices $f(U)$. This number is proportional to the
relative error on a smoothly varying observable, for a given statistics.
The top line is the standard choice.}
\vspace{-0.4cm}
\end{table}

To verify the improvement of the overlap with the desired measure, 
we generated ensembles of 200 
configurations ($\beta=4.8, n_f=2, ma=0.025, 4^4$ lattice), 
using the standard method ($\mu=0$) and using method A, 
i.e. at isospin $\mu_I \neq 0$.
Reweighting was then performed on each ensemble to obtain results at
isoscalar chemical potential $\mu=\mu_I$. 
Fluctuations of the reweighting factor were measured by
the numerator of Eq.(\ref{sd}). 
Table~1 shows that, for small $\mu$, the standard method (1rst line) 
gives larger fluctuations, i.e. a poorer sampling, than method A (2nd line).


Still, in either method the factor $\cos \theta$ must be computed for each 
configuration in the ensemble. For small $\mu$, it is possible to
approximate $\theta$ by its truncated Taylor expansion:
\bea
\theta & = & \frac{n_f}{4} {\rm Im~Tr}~\log M(\mu) = \bar\theta + {\cal O}(\mu^3
), \nonumber \\
\bar\theta & \equiv & \left.\frac{n_f}{4}~\mu~{\rm Im~Tr}~M^{-1}\dot{M}\right|_{\mu=0}
\label{Taylor}
\eea
where $\dis \dot{M}\equiv \frac{\partial M}{\partial \mu}$.
The advantage is that the trace can then be estimated using $n$ noise vectors $\eta_i$ as
\be
{\rm Tr}~M^{-1}\dot{M} \approx \frac1{n}\sum_i^n \eta^{\dagger}_i~M^{-1}\dot{M}~\eta_i.
\label{noise}
\ee
This approach has been used in \cite{Bielefeld}, with $n=10$, on a 
$16^3\times 4$ lattice. To test the quality of this approximation, we 
measured the exact phase $\theta$ and its linearized approximation $\bar\theta$ 
on each configuration of our isospin-$\mu$ gauge ensembles. 
The scatter plot Fig.~1 shows that the
correlation between $\theta$ and $\bar\theta$ is excellent at $\mu=0.1$
(i.e. $\mu/T=0.4$), and very poor at $\mu=0.3$. One may expect
this correlation to persist for larger volumes. Therefore, using
the linearized approximation $\bar\theta$ up to $\mu/T \sim 0.4$, as
advocated in \cite{Bielefeld}, appears justified. On the other hand, the
stochastic estimator of $\bar\theta$ Eq.(\ref{noise}) was extremely
noisy and required several thousand noise vectors.

\begin{figure}[tb]
\begin{center}
\includegraphics[height=6.5cm,width=7.5cm,clip]{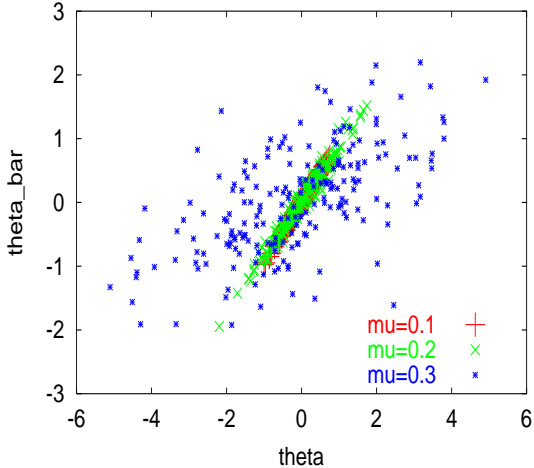} 
\end{center}
\vspace{-1.0cm}
\caption{The linearized phase $\bar\theta$ of the determinant vs the
exact phase $\theta$, for 3 values of $\mu$.}
\vspace{-0.8cm}
\end{figure}

\noindent $\bullet$ {\bf Method B: $f = e^{-\frac12 \bar\theta^2}$} \\
To approach the optimal $f = | \cos \theta |$ at small $\mu$ 
without computing higher derivatives in the Taylor expansion Eq.(\ref{Taylor}),
one can use 
$\cos\theta \approx 1 - \frac12 \bar\theta^2 \approx \exp(-\frac12 \bar\theta^2)$.
The additional term $S_{\bar\theta} = \frac12 \bar\theta^2$ in the action can be included
in an R-type algorithm~\cite{R-algo}, where the ${\rm Tr}\log$ is estimated
at each step using noise vectors. Here, two uncorrelated noise vectors are
necessary to estimate $\bar\theta^2$.
Maintaining the $\delta\tau^2$ accuracy of the R-algorithm
in the stepsize seems less obvious, but may be feasible.

To assess the advantages of method B over method A, we compare in Table~1
the standard deviation of the reweighting factor $\cos\theta / f$.
A large reduction is seen at small $\mu$.

\noindent $\bullet$ {\bf Method C: $f = | \cos\bar\theta |$} \\
To approach $| \cos \theta |$ even better at small $\mu$, one may consider
the choice $f = | \cos\bar\theta |$. As above, 
$\bar\theta$ can be estimated via Eq.(\ref{noise}). Then, an unbiased
estimator of $\cos\bar\theta$ can be formed by a stochastic Taylor expansion
as in~\cite{BK}, which can be used in the Monte Carlo update. 
As long as $\cos\bar\theta$ remains positive, i.e. for small $\mu$, 
an important gain is possible as seen in Table~1.

\section{CONCLUSIONS}
We have considered three choices of Monte Carlo actions which can be
used to obtain finite-$\mu$ results after reweighting. Compared to
the standard choice of simulating at $\mu=0$, they provide better 
statistical accuracy at small $\mu$ ($\mu/T \lap 0.5$), because the 
reweighting factor fluctuates less. In the case of Method A (isospin
chemical potential $\mu_I=\mu$), this improvement is achieved with
no computational overhead. Moreover, results at finite $\mu_I$ come for free.


\section*{ACKNOWLEDGEMENTS}
Simulations were performed on the SX-5 at RCNP, Osaka
University. T.T. is grateful to Electric Technology Research
Foundation of Chugoku for financial support. This work was supported
in part by Grant-in-Aid of the Ministry of Education, Culture, Sports,
Science and Technology (No.13740164).

\end{document}